\documentclass[a4paper, 11pt]{article}

\usepackage{amsmath}

\usepackage{graphicx}
\usepackage[space]{grffile}
\usepackage{tabulary}
\makeatletter

\usepackage[utf8]{inputenc}

\usepackage[T2A,T1]{fontenc}
\usepackage[english]{babel}
\usepackage{amsmath}
\usepackage{amsthm}
\usepackage{geometry}

\usepackage[round]{natbib}

\newcommand{\specialcell}[2][c]{%
  \begin{tabular}[#1]{@{}c@{}}#2\end{tabular}}

\title{Maximum likelihood estimation of parameters of spherical particle size distributions from profile size measurements and its application for small samples}
\author{Ekaterina Poliakova}

\begin{document}

\maketitle
\selectlanguage{english}
\begin{abstract}\small{
Microscopy research often requires recovering particle-size distributions in three dimensions from only a few (10–-200) profile measurements in the section. This problem is especially relevant for petrographic and mineralogical studies, where parametric assumptions are reasonable and finding distribution parameters from the microscopic study of small sections is essential. This paper deals with the specific case where particles are approximately spherical (i.e. Wicksell's problem). The paper presents a novel approximation of the probability density of spherical particle profile sizes. This approximation uses the actual non-smoothness of mineral particles rather than perfect spheres. The new approximation facilitates the numerically efficient use of the maximum likelihood method, a generally powerful method that provides the distribution parameter estimates of the minimal variance in most practical cases. The variance and bias of the estimates by the maximum likelihood method were compared numerically for several typical particle-size distributions with those by alternative parametric methods (method of moments and minimum distance estimation), and the maximum likelihood estimation was found to be preferable for both small and large samples. The maximum likelihood method, along with the suggested approximation, may also be used for selecting a model, for constructing narrow confidence intervals for distribution parameters using all the profiles without random sampling and for including the measurements of the profiles intersected by section boundaries. The utility of the approach is illustrated using an example from glacier ice petrography.}
\end{abstract}%

\section*{Introduction}
\begin{figure}[ht]
\begin{center}
\includegraphics[width=0.70\columnwidth]{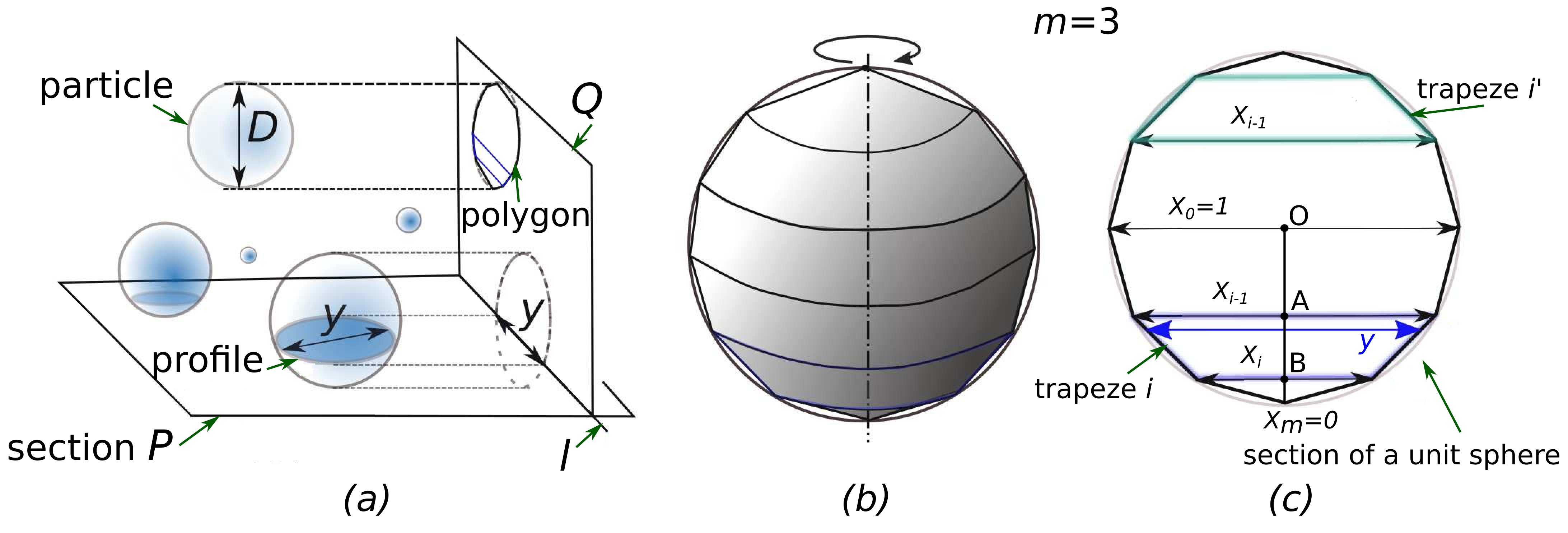}
\caption{Approximating spherical particles by solids of revolution: (A) The spherical particles, the notation and the $4m$-sided polygon on the plane $O$; (B) The approximating solid obtained by rotating the $4m$-sided polygon; (C) the denotations for the sizes of the trapezes, used for Eq. \ref{eq::0} and Eq. \ref{eq::4}.}
\label{fig::1}
\end{center}
\end{figure}

{\label{707961}}
When studying particle profiles in planar sections, their size and size distribution in three dimensions are often of interest as these differ from the sizes and distributions found in the sections (Fig. \ref{fig::1}, A). Generally, unbiased non-parametric methods are used to recover particle size distributions  \citep{cruz2017stereology}, which are well-developed for large sample sizes. The unbiasedness of estimates of such characteristics as mean radius, mean surface and mean volume is specially important in medicine and biology and is required by many journals when these characteristics are reported \citep{baddeley2004stereology}.
\par
However, the methods for small stereological sample study are much less developed, and, in these cases, minimising variance becomes more important than a slight bias. This is particularly relevant to Earth sciences, especially petrography, where a researcher, having samples which are impossible to increase (most typically - obtained from drilling), is interested in estimating a few quantitative characteristics of the samples and revealing global trends, patterns and irregularities, reflecting the difference in the evolution of some processes, e.g. in different geological structures. These processes lead to specific particle size distributions, so it is often reasonable to assume one or a few types of distribution, and it becomes sufficient to choose one out of several possible distributions and estimate the parameters. In this case, the bias, although deforming the trends, is of much less importance than, e.g., when comparing pairs of samples.

We encountered this type of problem when studying coarse-grained ice structures in an ice core from the World’s deepest glacier borehole at Vostok Station, East Antarctica \citep{lipenkov2016}. The ice core is of a relatively small size, and the sections typically contain only 10 -- 200 ice crystals, which makes it difficult to assess the accuracy of the ice structure characteristics measured. The spatial pattern of the distribution parameters is our main interest, because it provides information about the mechanisms influencing grain growth and recrystallisation, e.g. the pinning of grain boundaries by microparticles \citep{durand2006effect}. Another example of this problem which the author faced before, is the study of minor accessory minerals in granites of Chukchi region, North-East Russia \citep{alekseev2014evolution,alekseev2015accessory}. The properties of the monazite and zircon grains in that region are determined by the rare metal ore forming process, and it would be easier to predict the placement of ore deposits if the size of the mineral grains could be measured. However, the minerals are present in small amounts, and no more than 50 grains may be measured in a typical petrographic  thin section; so, though it is desirable, no such prediction has ever been made. 

Even for parameter estimation, non-parametric stereological methods are typically applied \citep[e.g.][]{gulbin2008estimation}. The variance of such estimates is well studied both theoretically and numerically \citep[ch. 13]{baddeley2004stereology}, and, when samples are too small for non-parametric methods, no alternative methods are used at all, though qualitative characteristics are applied, such as 'coarse-grained fabric'. Although parametric statistics is well-developed and provides more precise estimates when parametric assumptions are true, it is still rarely used in stereology. For spherical particle profiles, however, the moments are known, and the method of moments (MoM) is sometimes used \citep{goldsmith1967calculation}. The stereological use of minimum distance estimation (MDE) has also been recently described \citep{depriester2019resolution}. Still, the maximum likelihood (ML) method is 'the most popular technique for deriving estimators' \citep[p. 315]{casella2002statistical}, providing point estimators that are asymptotically unbiased and of lowest variance under mild assumptions and for large sample sizes. It is also a flexible and powerful instrument for estimating intervals and testing hypotheses \citep{lehmann2006testing}. ML has become increasingly popular \citep{schweder2016confidence} along with the development of computational methods. However, the stereological use of ML remains only rare \citep[e.g.]{hobolth2002note, keiding1972maximum}, and, in practice, it is typically applied to grouped data and empirical densities \citep{gulbin2008estimation}.

One of limitations of the parametric methods is that the estimates typically have a little bias, even when all assumptions hold, while the commonly applied methods follow demands from the biological sciences which include unbiasedness. However, the strict unbiasedness seldom applies to the geosciences \citep{gulbin2008estimation,lopez2016extension,durand2006ice}, and to some material studies \citep{depriester2019resolution}, where it is traditional to approximate convex non-spherical particles of irregular shape via spheres, estimating the conventional 'effective diameter' which includes the bias.  Another problem is the complicated numerical calculation of the probability density of the particle profiles’ probability density, which involves an improper integral even when the particles are spherical \citep{wicksell1925corpuscle}. This is especially relevant to ML and MDE: a single MDE estimation takes minutes even on a modern computer \citep{depriester2019resolution}. Finally, imprecision is typically unclear in cases where assumptions about particle shape and distribution type hold only approximately.

Following the tradition of the spherical approximation of particles, we will provide a useful approximation for the probability density of particle size, relying on actual non-spherical particle shape. Further, we will compare bias and variance for ML, MoM and MDE estimates of parameters (mean radius, shape and scale) of Weibull and log-normal size distributions (as is most usual in petrology) and positive normal distribution parameters (as recently used to present the applicability of MDE). We will then illustrate, on real bounded samples, that the method allows use to choose the type of distribution and ensure that the convex particles are approximately spherical. We will also compare interval estimates obtained by all these methods in several ways, applying random sampling and considering the limited sample sizes by censoring and weighting the likelihood.
\par\null
\section*{Methods and materials}
\subsection*{Approximation for the probability density of spherical particle profile sizes, given the probability density of sample sizes}\label{second-level-heading}

We approximate the probability density $g(y)$ of spherical particle profile sizes $y$, given the probability density of sample sizes as follows. 
We consider spherical particles (Fig. \ref{fig::1}) with centres that are uniformly distributed in a very large piece of volume and the diameters $D$ having some probability distribution with cumulative distribution function $F$ and probability density $f$.\par 
We call the section plane $P$ (Fig. \ref{fig::1}, a). We choose an arbitrary line $l$ on $P$ and draw a plane $Q$, which is orthogonal to $P$ and contains $l$. We project all the particles orthogonally onto $Q$. Their projections are of the same diameter as the particles; a particle is intersected by $P$ if and only if its projection is intersected by $l$. We choose a positive integer $m$ and assign to any particle a regular $4m$-sided polygon inscribed into the particle's projection and oriented in such a way that its longest diagonal is perpendicular to $l$ (and, hence, perpendicular to $P$). \par
 We will first introduce a biased intermediate approximation $g^*$ for the density of the diameters $y$ of the profiles of the particles. This density will be the density $g^*(y_s)$ of the profiles $y_s$ of the solids. Thereafter, we will obtain the main resulting approximation $g(y)$ via scaling transformation. To derive $g^*(y_s)$, we approximate any spherical particle by a solid of revolution obtained by rotating the $4m$-sided polygon around one of its diagonals that is perpendicular to $P$ (and moving back the solid so that its centre coincides with the centre of the particle  in Fig. \ref{fig::1}, b). \par

All the profiles of the solids are circles.  The $y_s$  have the same distribution as the lengths of the profiles of the $4m$-sided regular polygons intersected by $l$. The diameters $D$ of the intersected particles follow the size-weighted distribution \citep[p. 34]{baddeley2004stereology} with density 
\begin{equation}
\phi(t)=\frac{tf(t)}{E(D)},
\end{equation}
where $E(D)$ is the expected value of the particle diameter. Each of the $4m$-sided polygons is a union of $2m$ trapezia with bases parallel to $l$. We denote by $p_i$ the probability that given that a particle is intersected, the corresponding profile of the polygon  belongs to either the $i$-th trapezium or the trapezium which is congruent to  the $i$-th trapezium (denoted as $i'$-th trapezium in Fig. \ref{fig::1}, c). The $p_i$ is proportional to the  height of the $i$-th trapezium. Denoting the particle centre by $O$ and the distances from $O$ to the trapezium bases by $OA$ and $OB$, 
\begin{equation}
p_i=\frac{2AB}{D}, 
\label{eq::1explaining}
\end{equation}
and from the geometry of the regular polygon we calculate
\begin{equation}
p_i= sin\frac{i\pi}{2m}-sin\frac{(i-1)\pi}{2m},\quad for\quad i=1,...,m.
\label{eq::1}
\end{equation}

When a particle diameter $D$ is fixed  to equal some number $t$ and the number of the intersected trapeze is given, the distances $h$ from particle centres to the intersections between diameters of the profiles and the axis of rotation are distributed uniformly between lengths of $OA$ and $OB$, because of the initial assumption on the uniform distribution of the centres in the volume and the fixed $D=t$. Hence the diameters of the profiles are also distributed uniformly between $tx_{i-1}$ and $tx_{i}$ where the dimensionless constants $x_{i-1}$ and $x_{i}$ are bases of the $i$-th trapezium which is inscribed in the sphere with the unit diameter (Fig. \ref{fig::1}, c). The  $x_i$ are computed from the geometry of the regular polygon and equal:
\begin{equation}
x_i=cos\frac{i\pi}{2m},\quad for\quad i=0,...,m.
\label{eq::2}
\end{equation}
The diameters of the profiles of the solid have density \begin{equation}
g^*(y_s|D=t,i)=\begin{cases}\frac{1}{t\cdot(x_{i-1}-x_{i})}, y_s\in(tx_{i},tx_{i-1})\\ 0,\quad otherwise .\end{cases}
\end{equation}
When a particle diameter $D$ is fixed to equal $t$ but a trapezium may be arbitrary, the density of the profiles is \begin{equation}
g^*(y_s|D=t)=\sum_{i=1}^m p_i g^*(y_s|D=t,i),
\end{equation}
where $p_i$ is the probability to belong to the $i$-th trapeze, computed by the Eq. \ref {eq::1}.
When $D$ is not fixed, we observe that the condition $ y_s\in(Dx_{i},Dx_{i-1})$ is equivalent to $ D\in(\frac{y_s}{x_{i-1}},\frac{y_s}{x_{i}})$, and therefore the probability density of the profiles becomes
\begin{equation}g^*(y_s)=\int_{y_s}^\infty g^*(y_s|D=t) \frac{tf(t)}{E(D)}dt = \frac{1}{E(D)}\sum_{i=1}^m\frac{p_i\Big(F\big(\frac{y_s}{x_{i}}\big)-F\big(\frac{y_s}{x_{i-1}}\big)\Big)}{x_{i-1}-x_{i}}.
\label{eq::0}
\end{equation}
The coefficients  $p_i$ and $x_i$ are computed from Eqs. \ref{eq::1} and \ref{eq::2}.\par
This intermediate approximation $g^*$ is applicable but inexact if the number of summands $m$ is small, because the approximating profiles of the polygon are shorter than in reality and hence its graph is deformed as compared to the graph of the true probability density. In order to reduce the bias of the mean diameter estimates, the approximation may be improved by applying a scaled transformation, so that the mean of the approximation $g(x)$ is scaled to equal the mean of the actual density. For any fixed $D=t$, $E(y_s)$ in the intermediate approximation equals 
 \begin{equation}
\begin{aligned}
E(y_s)=\int_{y_s=0}^t y_sg^*(y_s|D=t)dy_s=\sum_{i=1}^m\frac{x_{i-1}+x_{i}}{2}tp_i=\\=\sum_{i=1}^m\frac{Area(trapeze_i)+Area(trapeze_{i'})}{t}=\frac{Area(regular\:polygon)}{t}=\frac {m t}{2} sin\frac{\pi}{2m},
\end{aligned}
\label{eq::not_used}
\end{equation}
but 
 \begin{equation}E(y)=\frac{Area(profile)}{t}=\frac{\pi t}{4}
 \label{eq::not_used2}
\end{equation}
actually (as the sum for infinitely many narrow trapezia). So we choose the scaling coefficient
\begin{equation}
   a=\frac{\pi}{2m\: sin\frac{\pi}{2m}}. 
   \label{eq::3}
\end{equation} The final form of approximation becomes 
\begin{equation}g(y)=\frac{1}{aE(D)}\sum_{i=1}^m\frac{p_i\big(F(\frac{y}{ax_{i}})-F(\frac{y}{ax_{i-1}})\big)}{x_{i-1}-x_{i}},
\label{eq::4}
\end{equation}
where $p_i$, $x_i$,  and $a$ are defined by Eqs. \ref{eq::1}, \ref{eq::2}, \ref{eq::3} respectively, and $E(D)$ is computed from the known distribution of three-dimensional particles and depends on the parameters.\par

\par\null
\subsection*{Materials and numerical computations}

All the following numerical computations were completed in  \verb"R", except for the control density computation (first line in Table \ref{tab::1}, done by using the function \verb"integral" in \verb"MATLAB" (online version 20.0 R2020a [9.8.0]). We focus mainly on log-normal and Weibull particle size distribution that are most common for petrographical studies, and we also consider positive normal particle size distribution (i.e. truncated normal distribution with positive mode, bounded from below at 0), because it was mentioned as typical for a sample similar to ours by \citep{depriester2019resolution}. All these distributions are characterized by shape and scale parameter:  $\mu$ and $\sigma$, $\lambda$ and $k$, $\mu$ and $\sigma$ respectively,  and we denote the specific distributions as: 'log-normal$(\mu,\sigma)$', 'Weibull$(\lambda,k)$', and 'positive normal$(\mu,\sigma)$'.

The ML estimates were computed via numerical maximization of the likelihood. Following
 \citeauthor{casella2002statistical}  (\citeyear{casella2002statistical}, p.~290), 
or a more elementary description \citep[p.~284]{larsen2013introduction}, the \textit{likelihood} for each individual profile measurement $x_i$ is a probability density computed in this measurement, as a function of the unknown parameters. When profile sizes are independent, the likelihood of the sample is a product of the likelihoods for individual measurements, treated as a function of the parameters. The ML estimates are the values of the parameters that maximise the likelihood, given the measurements. The densities involved were approximated by Eq. \ref{eq::4}. These estimates are called further 'ordinary' ML estimates, when contrasted with censored ML and weighted ML, described below.

MoM estimates were computed using the known \citep[p. 37]{baddeley2004stereology} expressions for diameter-weighted distributions of sphere sizes, which provide the expected values for the first and second degrees of profile diameters. These expressions are set equal to the actually observed mean values of the first and second degree of profile diameters and solved with respect to parameters. Denoting $\overline{Y}$ for the mean diameter of the profiles, $S_Y
^2$ for the sample variance of the profiles, $\overline{Y^2}$ for the mean square of the profiles and hat $\:\hat{}\:$ for the estimates as functions of data, we used the following expressions for the log-normal distribution:
\begin{equation}
\begin{cases}
\hat{\sigma^2}=max\big(log(S_Y^2)-2log(\overline{Y})+2log(\pi)-3log(2)+log(3),0\big)\\
\hat{\mu}=log(\overline{Y})-log(\pi)+log(2)-\frac{3}{2}\hat{\sigma^2}.
\end{cases}
\label{eq::mom1}    
\end{equation}
We also used the following expressions for the Weibull distribution:
\begin{equation}
\begin{cases}
\hat{k}=argmin\Big|\frac{32\Gamma(1+3/k)\Gamma(1+1/k)}{3\pi^2(\Gamma(1+3/k))^2}-\frac{\overline{Y^2}}{\overline{Y}}\Big|\\
\hat{\lambda}=\frac{4\overline{Y}\Gamma(1+1/\hat{k})}{\pi\Gamma(1+2/\hat{k})},
\end{cases}
\label{eq::mom1a}    
\end{equation}
where 'argmin' denotes that the $\big|\frac{32\Gamma(1+3/k)\Gamma(1+1/k)}{3\pi^2(\Gamma(1+3/k))^2}-\frac{\overline{Y^2}}{\overline{Y}}\big|$ was minimised numerically to be as close to zero as possible, and $\Gamma$ is the gamma-function.

The corresponding estimates for the positive normal distribution were computed similarly, by minimising the sum of the squared differences between the observed and theoretical values of the means and variances, using the expressions for the moments of truncated normal distribution from \citep{horrace2015moments}. 
MDE-estimates were computed in \verb"R", following the instructions of \citet{depriester2019resolution}, except when computing the inolved probability densities. The probability densities used in the MDEs were computed directly from Wicksell's equation \citep{wicksell1925corpuscle}. This is accomplished in R via the trapezoidal rule, with a varying, non-uniform grid, to use open software and to control the numerical efficiency (which has increased by 4 -- 20 times compared to \verb"MATLAB"). 

The particle profile diameters were simulated as sections of loose spheres. The diameters of the spheres intersecting the section were sampled from diameter-weighted distribution with given parameters and have been element-wise multiplied with the diameters for random profiles of unit spheres. When using bootstrap to estimate confidence ranges for non-random samples in limited sections (Table \ref{tab::1}), we added the weights computed by Eq. \ref{eq::w}. For plotting Figs \ref{fig::3} -- \ref{fig::3a}, 5000 simulations have been completed for each point.

\begin{figure}[h!]
\begin{center}
\includegraphics[width=0.40\columnwidth
]{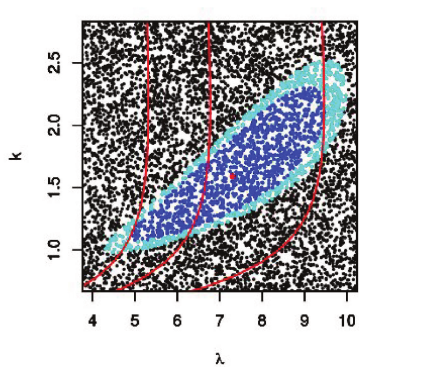}
\caption{Numerically estimating the confidence ranges for the Weibull distribution via the likelihood ratio method. The sample 3434-7, of $n=54$ profiles. The points $(\lambda, k)$ were sampled randomly and the likelihood $L$ was computed in each point. The set of values $(\lambda, k)$ where $-2logL>\Lambda_2(0.95,n)=6.25$ is the 95\% confidence range for the two-dimensional parameter and is shown in cyan. The three red curves $\lambda=\frac{D}{\Gamma(1+1/k)}$, where $D$ took values of $4.85$, $6$, and $8.56$ are the set of values $(\lambda, k)$ corresponding to these three values of D. The set of values $(\lambda, k)$ where $L<\Lambda_1(0.95,n)=4.0$ corresponds to the 95\% range of the one-dimensional parameter $D$ and is shown in blue. When $(\lambda, k)$ take values in the blue region, the $D$ takes values in the range $4.85..8.56$. The red dot is the maximum likelihood estimate $(\hat{\lambda}, \hat{k})$}
\label{fig::2}
\end{center}
\end{figure}

Error estimation by parametric bootstrap was completed as in \citet[ch. 9.2]{givens2012computational} and included bias correction. 25000 simulations were conducted for each ML and MoM estimate of mean radius in a real sample, in versions with random sampling and sampling all inner profiles not intersected by boundaries. For weighted likelihood, only 2000 simulations were completed for each sample, due to the lengthy computation of the normalising coefficient for the weighted density.
The confidence intervals for ML estimates were constucted using the 'Wilks theorem'. For small samples and  $k-$dimensional parameters, this theorem provides
\begin{equation}
2(logL-log\hat{L_0})<\Lambda_k(p,n)\label{eq::ci},
\end{equation} where $\hat{L_0}$ is the supremum of the likelihood, given the data and the assumed model, $L$ is the supremum of the likelihood over any possible parameters,  and where the critical value $\Lambda_2(p,n)$ depends on the sample size $n$ and $\Lambda_2(p,n) \approx \chi_{p,k}^2$, and $\chi_{p,k}^2$ is $p$-quantile of chi-square distribution with $k$ degrees of freedom. The  $p$-confidence ranges for the parameters contain the values satisfying this inequality. In order to estimate these ranges, we simulated 50000 possible values of $\lambda$ and $k$-parameters (we sampled them from two positive normal distributions with modes at the ML estimates ($\hat{\lambda},\hat{k}$) and standard deviations approximately five times as large as found from simulations for Figs \ref{fig::3} -- \ref{fig::3a}). The possible true likelihood $\hat{L_0}$ was computed in all the parameter points as a function of data, $\lambda$ and $k$, while the $\hat{L}$ was the actually observed maximum likelihood for the data.   To obtain the critical value, the actual 90- and 95\%-quantiles of $2(logL-log\hat{L_0})$ were computed from 30000 simulations for seven values of the shape parameter and three sample sizes (20, 30 and 50) in each of the distributions. Then, for the actual small sample sizes, $\Lambda_2(p,n)$ was roughly estimated numerically as: $\chi_{0.97,2}^2$ for $n=18..19$, $\chi_{0.96,2}^2$ for $n=20..50$, $\chi_{0.955,2}^2$ for $n>50$). The ML estimate for the parameter 'mean diameter' is $\hat{\lambda}(1+1/\hat{k})$, and the $p$-confidence ranges were taken to be the ranges of  $\hat{\lambda}(1+1/\hat{k})$ such that $L(data, \lambda, k)<\Lambda_1(p,n)$ (where $\Lambda_1(p,n)$  was estimated numerically same as above, but for one degree of freedom). 
 This way of numerical computing the confidence intervals is illustrated in Fig. \ref{fig::2}. 

When samples are very small, it may be preferable to use all available measurements. For this purpose, we calculated maximum censored likelihood estimates and maximum weighted likelihood estimates. Censored likelihood includes boundary intersected profiles of size $y_j$, which are usually not measured. However, measuring part of a grain profile intersected by the section boundary means that the whole profile is larger than the measurement, and yields a \textit{censored} measurement occuring with a probability $P_j=1-G(y_j)$, where $G$ is the cummulative distribution function of profile sizes. We compute the likelihood as a product of the probability densities of the inner profile measurements and the probabilities of the censored measurements, for which Wilks’s theorem also holds \citep{schweder2016confidence, borgan1984maximum}. The value of $G(y_j)$ was computed by numerically integrating of the probability densities. The confidence intervals for the maximum censored likelihood estimates were computed using same critical values as chosen for ordinary likelihood, although the validity of the latter was checked for only 5 combinations of sample sizes and shape parameters. Weighted likelihood includes stochastic weights which are proportional to the possibility of sampling \citep{wang2001maximum}. When all profiles not intersected by a section boundary are sampled, this is equivalent to the fact that the squares circumscribed around the circular profiles and having sides $y_j$ parallel to the section sides, have all four vertices inside the rectangular sections with sides $s_1$ and $s_2$. Therefore we used in the cases of limited sections the weights 
\begin{equation}
w_j=(s_1-y_j)(s_2-y_j),
\label{eq::w}
\end{equation}
such that the weighted density $g_w(y_j)$ is proportional to $w_jg(y_j)$, where $y_j<min(s_1, s_2)$ and the coefficient of proportionality can be calculated numerically for each combination of parameters.

As the variance of non-parametric estimates was estimated many times before, we only compare the parametric estimators with the estimator based on Saltykov method and described by \citet{gulbin2008estimation}, as a relevant to similar petrographic problems. Following \citet{gulbin2008estimation}, we chose log-normal($ln(1), 0.5$) distribution and simulated samples of 200 and 2000 profiles. The simulation conditions differed from those of \citep{gulbin2008estimation} where 30000 spheres in a limited volume should have been intersected, while we had no assumptions on maximal possible diameter.

\begin{figure}[h!]
   \begin{minipage}[b]{0.32\textwidth}
 \scalebox{0.24}{\includegraphics
 {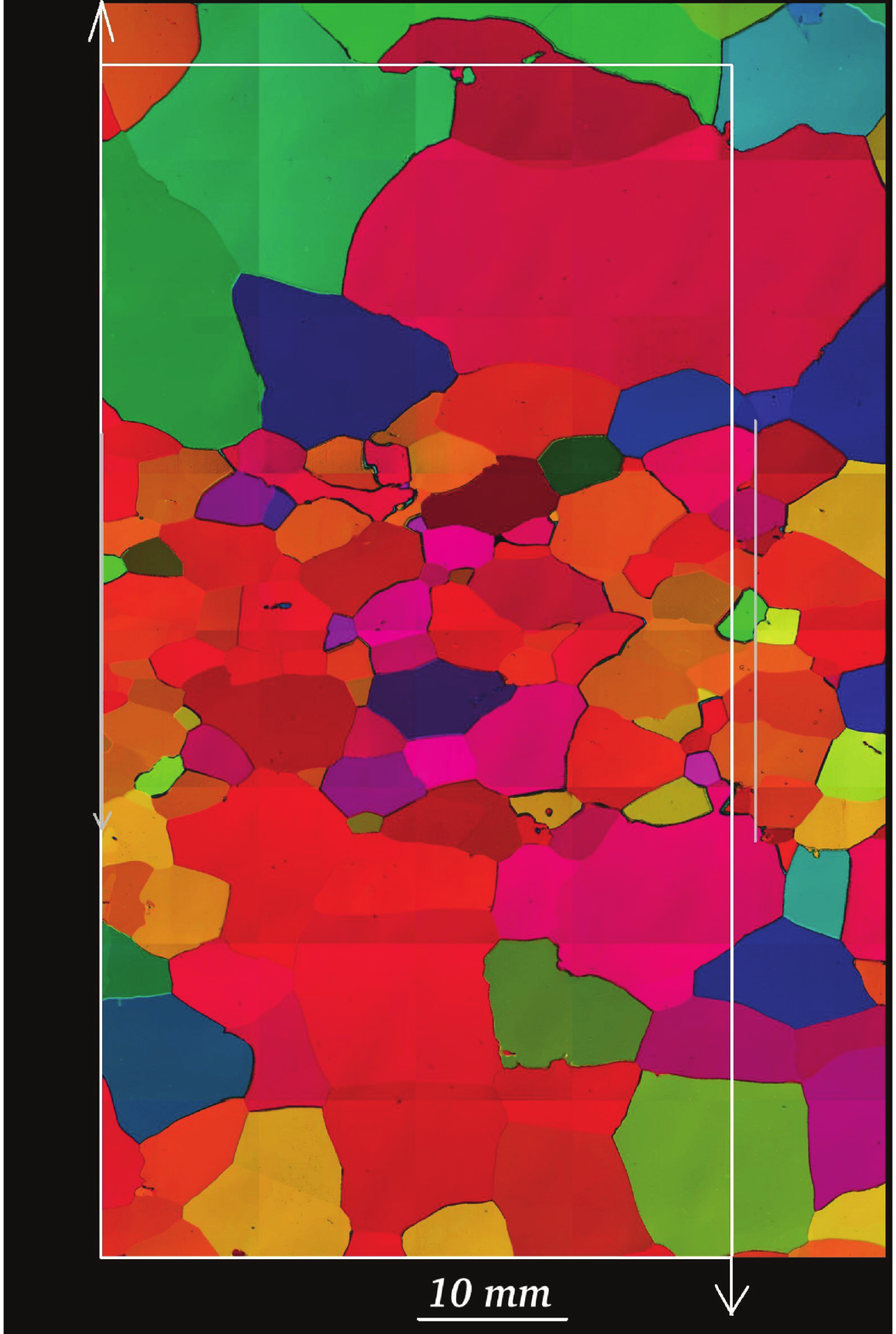}}
  {(A)}
  \end{minipage}
  \hfill
  \begin{minipage}[b]{0.32\textwidth}
 \scalebox{0.24}{\includegraphics
 {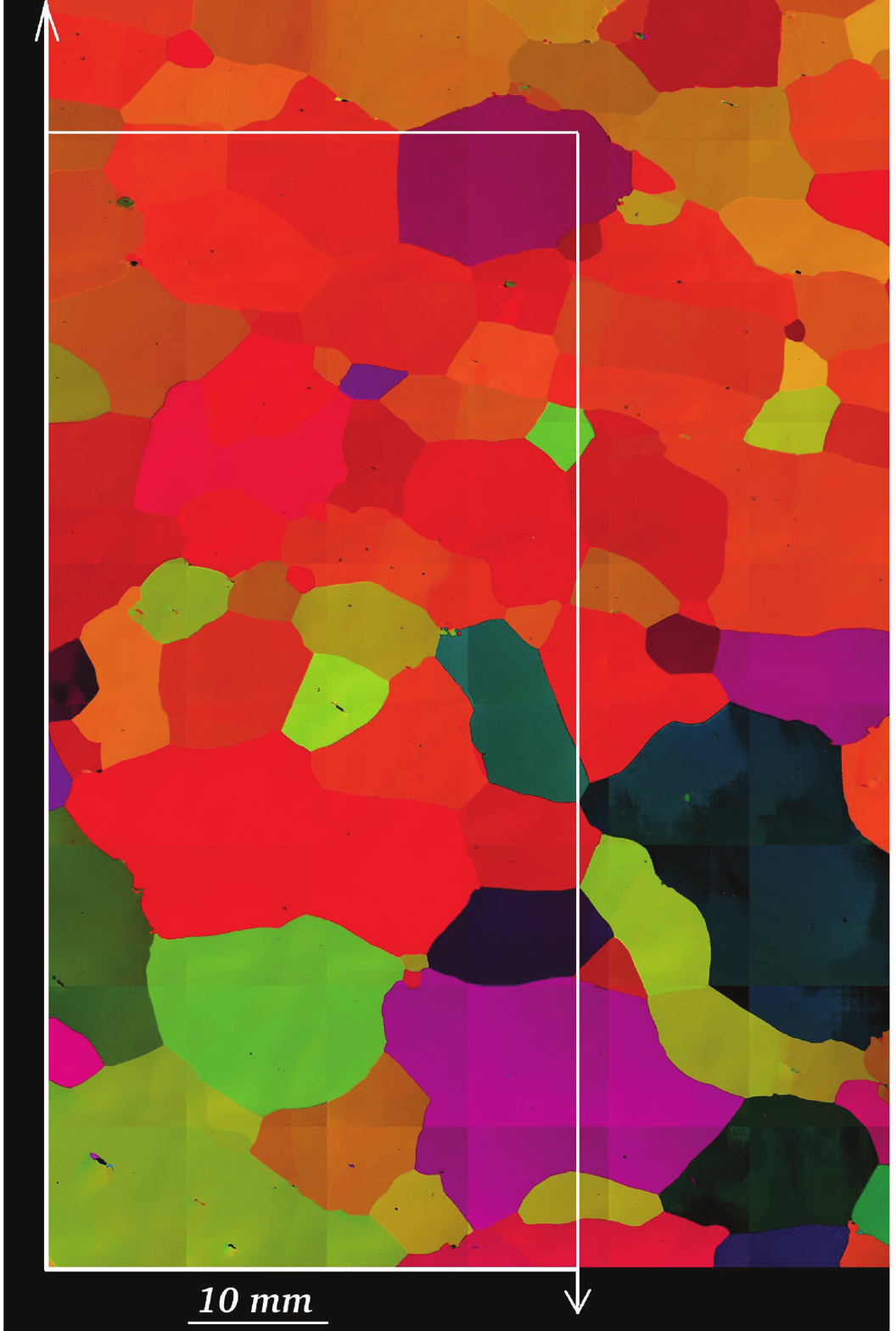}}
   {(B)}
  \end{minipage}
   \hfill
  \begin{minipage}[b]{0.32\textwidth}
 \scalebox{0.25}{\includegraphics
 {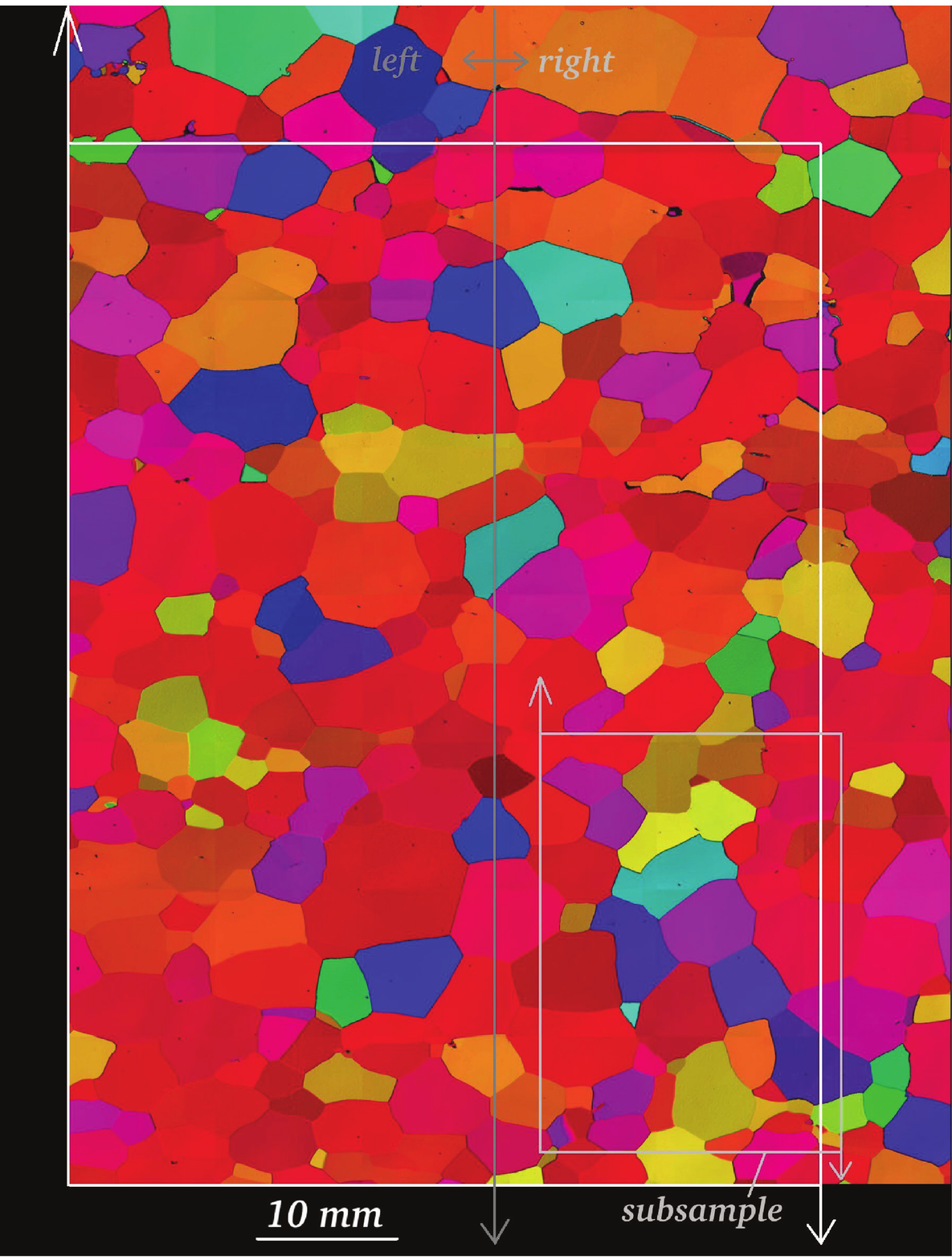}}
   {(C)}
  \end{minipage}  
   \caption{The thin sections of glacier ice studied (transmitted light, crossed polarisers): (A) a layer of fine-grained ice in a matrix coarse-grained ice, 3437.5 m (samples '3437-6 layer' and '3437-6 matrix'), (B) middle-grained ice, 3434.5 m (sample '3434-7'), (C) fine-grained ice, 3438.2 m (samples '3438-3' and '3438-3 subsample'). The lines depict boundaries of sampling area by tiling rule}
   \label{fig::b2}
\end{figure}

ML, MoM and MDE were applied to three pictures of glacier ice (Fig. \ref{fig::b2}), taken with an Automatic Ice Texture Analyser (AITA, Russel-Head Instrument, Templestowe, Australia \citep{Wilson2003}). These are thin sections of ice fabric in polarised transmitted light -- crossed polarisers, and the pictures have been the samples studied. They belong to a series of 208 micrographs which were taken at an interval of 25 m, along the world’s deepest glacier borehole at Vostok Station, East Antarctica, from a depth range of 3429 - 3452 m. The chosen samples represent the variety of the ice fabric at this depth. The sample in Fig. \ref{fig::b2}, (A) represents  a layer of fine-grained ice '3737-6 layer' in a coarse-grained ice matrix '3737-6 matrix'. Sample '3434-7' (Fig. \ref{fig::b2}, B) represents middle-grained ice, and sample '3438-3' (Fig. \ref{fig::b2}, C) represents fine-grained ice.  To investigate the efficiency of the methods, parameters are also estimated in a small subsample '3438-3 subsample', shown in the lower right part of Fig. \ref{fig::b2}, (C). Besides, in order to roughly observe the variability of the mean volume weighted volume, it was estimated in left and the right parts of '3438-3' separately. When computing weighted likelihood, all profiles not intersected by a section boundary were sampled. When computing censored likelihood, all profiles were sampled. Otherwise, the tiling rule \citep{gundersen1977notes} with boundary lines including the left and low sides of the section was used for sampling. The grain boundaries were drawn in \verb"MATLAB", ImageJ and refined in Adobe Photoshop CS3, and the areas of the profiles were measured in Adobe Photoshop CS3. All grains are single crystals, and, along with the profile areas, the orientation of the sixth order symmetry axis (i.e. $c$-axis) was measured as well in each grain. The colour in the pictures depends mainly on the inclination of the $c$-axes, but even profiles with similar $c$-axis-inclinations exhibited unequal c-axis azimuths (differing by at least 2–3 degrees). We observed that, in all the sections, non-convex areas of ice of similar orientation are decomposed into the disoriented convex profiles of blocks with clear boundaries. Hence, we assumed that the same approximate convexity in three dimensions when treating each block as a separate grain, although blocks may sometimes have slightly deeper areas on the surface, similar to those observed on profile boundaries.
When studying the grain size distributions in real samples, the model was as follows. The grains were approximated as spherical particles, with profile diameters $Y=\sqrt{S/4\pi}$ where $S$ was the grain profile area. We assumed that the set of profile sizes in sections was distributed as if the profile sizes were independent. (Dependence between adjacent grains in real samples could occur e.g. due to recrystallisation processes). If this assumption does not hold, or if large samples should be studied more thoroughly, the parametric methods may still be used, but in more complex ways \citep[e.g.][]{cressie2015statistics, ohser20093d} that are beyond this work. The diameters of the grains were assumed to be either Weibull$(\lambda,k)$ or log-normally$(\mu,\sigma)$ distributed.

When checking whether the particle volume corresponds to the assumptions, this model was refined. It considered that the glacier was deformed due of its heavy mass, and the ice grains were vertically compressed, a fact observed in all samples. In the refined model, we assumed that the particles were ellipsoidal with two equal horizontal axes $D_x$, $D_y$ and a third vertical axis $D_z=k\cdot R_x$. The coefficient $k<1$ varied between the samples (because of differences in the recrystallisation processes) and was found via clear linear regression between the height and width of the profiles in each sample. The effective diameters of the profile $Y$ and the distribution of the effective diameters $D$ were as above and were estimated in the same manner with same results, but then $D=\sqrt{k}D_x=D_y/\sqrt{k}$.

The unbiased estimation of the volume weighted mean volume was completed according to \citep{karlsson1992new}. The deformation of the fabric was determined as follows. The $b$ probe points were scattered randomly within a rectangular counting frame (with $b$ ranging between 200 and 400), and horizontal line sections $l_i$ were drawn through these points, with endpoints on the boundaries of the sections on which the points fell. The volume of a k-times stretched particle was estimated as $\hat{V}=k\frac{\pi\sum_{i=1}^bl_i}{3b}$. Therefore, if the shape assumption was true, the volume computed from the parametric assumptions and spheres with a random effective diameter should have approximately equalled $k^{0.5}\hat{V}=k^{1.5}\frac{\pi\sum_{i=1}^bl_i}{3b}$.
\par\null
\section*{Results}
\subsection*{Precision of the approximation}
We compared our approximation of the probability density of the diameters, for different number of summands $m$ and with the density computed directly from Wicksell's equation by numerical integration in \verb"MATLAB".  Examples of the correspondence between the results are presented in Table \ref{tab::1}. 
\begin{table}[h!]
\centering
\normalsize\begin{tabulary}{1.0\textwidth}{CCCCC}
Methods of Computation& Weibull
($1,0.9$) & Weibull
($1,1.2$) & Log-normal ($0,0.7$)& Positive normal ($3.876,2.816$) \\
\hline
Eq. (\ref{eq::4}), $m=8$ & 0.37034  &      0.50279  &      0.47966  &      0.07421 \\
Eq. (\ref{eq::4}), $m=15$  & 0.37169  &      0.50450   &      0.48126  &      0.07662\\
Eq. (\ref{eq::4}), $m=100$ & 0.37223   &       0.50515  &       0.48189  &       0.07655 \\
Eq. (\ref{eq::4}), $m=1000$ &0.372242  &       0.505167  &   0.481899      &   0.076559\\
Wicksell's equation, integration by function ''integral'' in MATLAB  &  0.372242 &   0.505167  & 0.481900 & 0.076559 \\
\end{tabulary}
\caption{Values of the probability density for the profile diameter $Y=1$ by different approximations and for different distributions}
\label{tab::1}
\end{table}
\par\null
\subsection*{ML compared with other methods}
\begin{figure}[h!]
\begin{center}
\includegraphics[width=\columnwidth
]{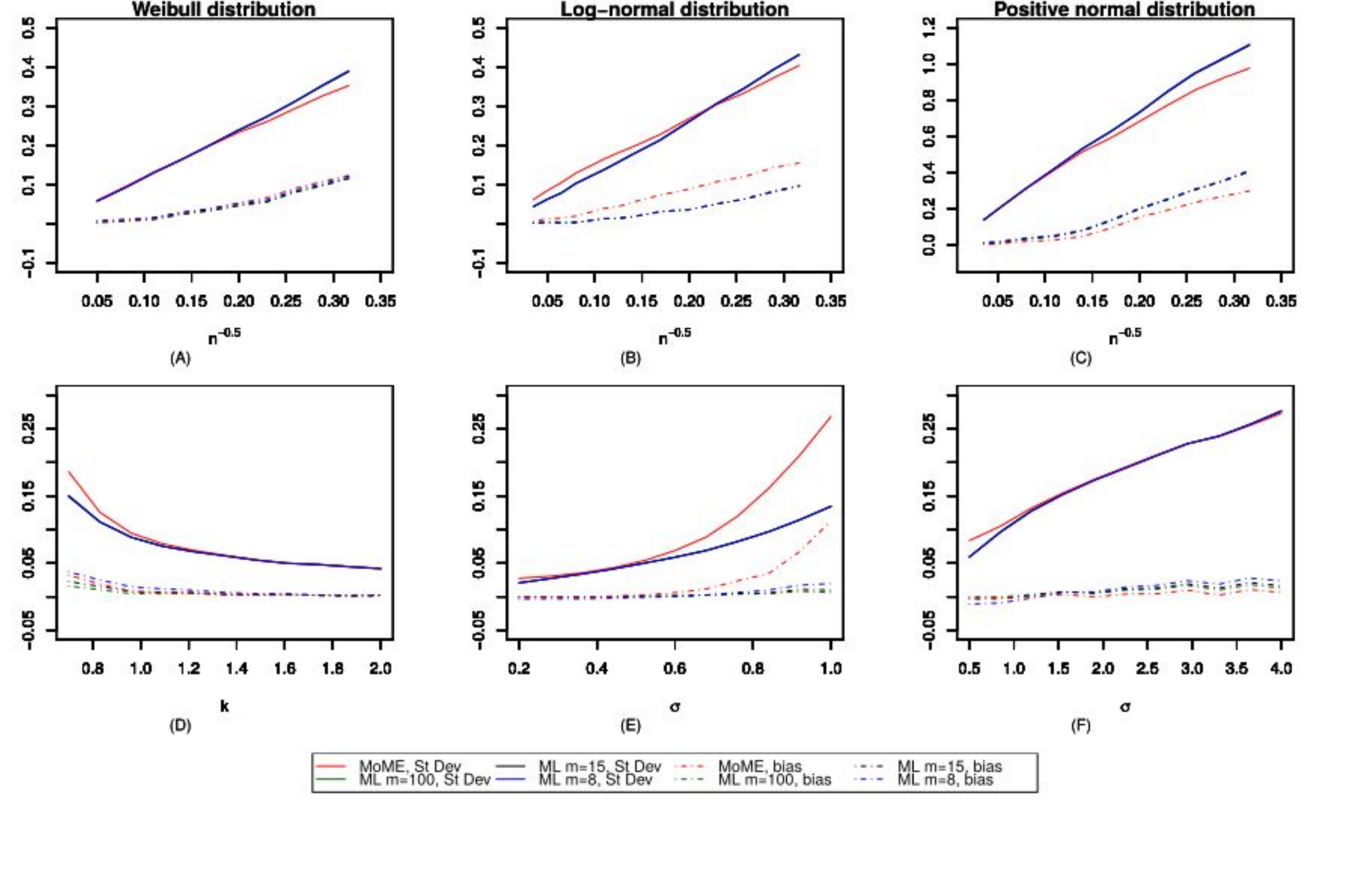}
\caption{The standard deviation and the bias of mean diameter estimates, as functions of $1/\sqrt{n}$ (where $n$ is the sample size) and of shape parameter. (A) Weibull distribution: $\lambda=1$, $k=1.2$, $1/\sqrt{n}$ varies; (B) log-normal distribution: $\mu=0$, $\sigma=0.7$, $1/\sqrt{n}$ varies; (C) positive normal distribution: $\mu=3$, $\sigma=3$, $1/\sqrt{n}$ varies; (D) Weibull distribution: $\lambda=1$, $n=300$, $k$ varies; (E) log-normal distribution: $\mu=0$, $n=300$, $\sigma$ varies; (F) positive normal distribution: $\mu=3$, $n=300$, $\sigma$ varies.}
\label{fig::3}
\end{center}
\end{figure}

\begin{figure}[b!]
\begin{center}
\includegraphics[width=0.90\columnwidth
]{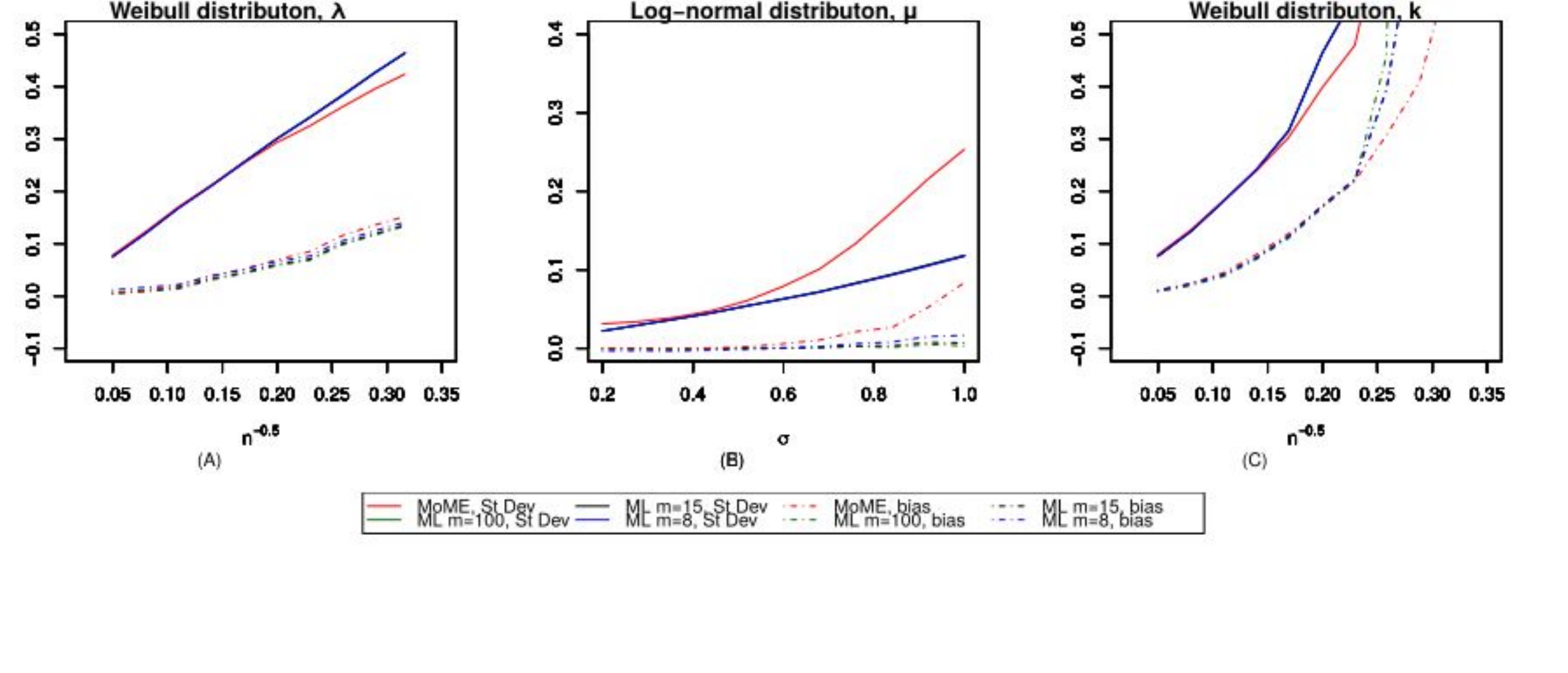}
\caption{The standard deviation and the bias of shape and scale parameter estimates in Weibull distribution as functions of $1/\sqrt{n}$ (where $n$ is the sample size) and of shape parameter, typical examples.  (A) Weibull distribution: $\lambda=1$, $k=1.2$, $1/\sqrt{n}$ varies; (B) log-normal distribution: $\mu=0$, $n=300$, $\sigma$ varies; (C) Weibull distribution: $\lambda=1$, $k=1.2$, $1/\sqrt{n}$ varies.}
\label{fig::3a}
\end{center}
\end{figure}

Mean diameter $E(D)$ is one of the most practical parameters, and it is relevant to all spherical particle size distributions. Fig. \ref{fig::3} presents typical dependencies of sample sizes $n$ and shape parameters, which we observed for the bias and standard deviation of ML and MoM estimates. Overall, for each shape parameter, there is a minimal sample size (17–70 profiles, depending on the distribution type and shape) under which MoM estimates have lower standard deviations than ML estimates. In the Weibull distribution, the bias of ML estimates is almost equal to that for MoM; in the log-normal distribution, the ML bias is much lower than the MoM bias; and, in the positive normal distribution, the ML bias is slightly larger than the MoM bias. Except for the log-normal distribution, where ML is clearly preferable because of the lower bias for $n\geq 10$ and lower variance for $n\geq 20$, the properties of the ML and MoM estimates are rather similar.
 Fig. \ref{fig::3a} shows three typical dependencies for the other parameters. The behaviours of these other shape and scale parameters, according to sample size and distribution shape, are qualitatively close to the behaviours observed for mean diameter. For the scale parameter estimate, the dependencies are almost identical. However, the shape parameter estimates are more imprecise (compared to those for mean diameter and scale) for very small sample sizes, and their properties exhibit more evident differences between MoM and ML methods. For MDE, among all estimates, both standard deviation and bias are observed to be much larger than those for ML and MoM. For sample sizes <50 profiles standard deviation and bias were beyond the axis limits in Figs  \ref{fig::3}--\ref{fig::3a}, and their estimation took unreasonably many days because of large frequency of totally wrong estimates which were far from the initial values in the minimisation and made computations too slow. Therefore, they are compared with ML and MoM estimates only in Table \ref{tab::1.5} for sample sizes 200 and 2000. 

Concerning the applicability of the approximation, in Figs \ref{fig::3}--\ref{fig::3a} we observe that the standard deviation of the estimates is almost equal, whether $m=8$ or $m=100$ summands in the approximation. The bias caused by too few summands is, in most cases, indistinguishable as well. However, for as few as eight summands in the approximation and for some shape parameters, the bias can reach 1/5 of the standard deviation. The difference between the bias for $m=100$ and $m=15$ is almost indistinguishable, hence we considered the estimates with $m=15$ be approximately equal to those corresponding $m=\infty$ and used $m=15$ studying the real samples further.  In addition, the approximation Eq. \ref{eq::4} works from two to three times faster than the equally exact density computation, integrating Wicksell's equation via the trapezoidal rule, and MLE using this approximation works six to 30 times faster than MDE when applied to the same samples. We also compared  the computation time for MDE for the sample from positive normal distribution described by \citet{depriester2019resolution} with the time cited as having been 120.5 s, and on the author's computer the computation was 1.2 times faster. MDE computation can be improved by approximating the densities involved in Eq. \ref{eq::4}, but, as this includes computing the cumulative distribution function by integrating densities numerically, it is not likely to be more efficient than using densities directly.

\begin{figure}[h!]
\begin{center}
\includegraphics[width=0.40\columnwidth]{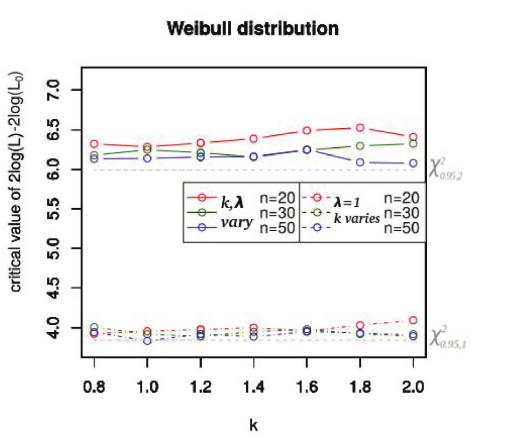}


\caption{The quantiles for the likelihood in the Weibull distribution as functions of shape parameter, for sample sizes 20, 30 and 50. Upper part: both parameter vary; lower part: $\lambda=1$, $k$ varyes. Each point is computed from 30000 simulations.}
\label{fig::30}
\end{center}
\end{figure}

An important fact for confidence interval construction is that the quantiles  of $2(log L-log\hat{L_0})$ are almost independent on the shape parameter not only for very large samples, but also for moderate sample sizes of 50 profiles and more. They are also very close to the asymptotic quantiles predicted by Eq. \ref{eq::ci}. For Weibull distribution, the quantiles of $2log L$ are almost stable from a sample size of 20 profiles (Fig. \ref{fig::30}). That facilitates evaluating the maximum possible value of $2(log L-log\hat{L_0})$, which should be taken to construct confidence intervals of  a chosen coverage. The 95\% and 90\% confidence ranges of both one- and two-dimensional Weibull parameter estimates may be chosen by selecting the parameters below the 96\% and 91\% quantiles of the chi-squared distribution (one degree of freedom for a one-dimensional parameter; two degrees of freedom for a two-dimensional parameter, i.e., when both $\lambda$ and $k$ may vary independently).  For sample sizes >50, the 95.5\% and 90.5\% quantiles can be taken as critical values.

 The comparison of ML, MoM and MDE with a non-parametric Saltykov-type estimation method is provided in Table \ref{tab::1.5} for a log-normal size distribution with typical \citep{gulbin2008estimation} parameters. We observe that the estimates by parametric methods are more than 10 times preciser than those by the approach including a non-parametric method. The ML has still lowest bias and lowest variance.

\begin{table}[h!]
\centering
\normalsize\begin{tabulary}{1.0\textwidth}{CCCCCC}
Sample size && ML & MoM  & MDE & Saltykov, q=20 \\
\hline
n = 200 &  bias& $2.5 \cdot 10^{-3}$  &  $4.2 \cdot 10^{-3}$  &  $-14.5\cdot 10^{-3}$  &  $228\cdot 10^{-3}$ \\
& st. dev.& $ 6.1 \cdot 10^{-2}$  &  $ 6.7 \cdot 10^{-2}$  &  $6.4\cdot 10^{-2}$  &  $66\cdot 10^{-2}$ \\
\hline
n = 2000  &  bias& $0.12\cdot 10^{-3}$ &   $1.0\cdot 10^{-3}$   &  $-5.3\cdot 10^{-3}$ &      $310\cdot 10^{-3}$ \\
& st. dev.& $1.9\cdot 10^{-2}$ &  $2.2 \cdot 10^{-2}$   &  $ 2.0\cdot 10^{-2}$ &     $71 \cdot 10^{-2}$ \\
\end{tabulary}
\caption{The biases and standard deviations (st. dev.) of the estimates of the median diameter of log-normally$(0,0.5)$-distributed spheres. The true value is 1. Each line corresponds 1000 simulated samples, resulting in 1000 estimates for each entry. ML estimates were computed using our approximation, MoM by using the known moments and \citep[p. 37]{baddeley2004stereology}, MDE by \citep{depriester2019resolution}, Saltykov method by \citep{gulbin2008estimation} using $q=20$ size classes.}
\label{tab::1.5}
\end{table}

\subsection*{Application to real ice samples}

\begin{figure}[h!]
\begin{center}
\includegraphics[width=0.4\columnwidth]{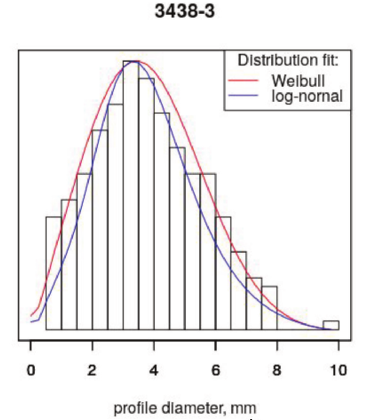}
\caption{The fit between the estimated and observed probability densities for the sample 3438-3.}
\label{fig::5}
\end{center}
\end{figure}

First, we tested whether the grain distributions belong to the log-normal distribution, which occurs in such ice in normal grain growth \citep{durand2006ice}, or to the Weibull distribution, which is is typical for grain fractioning processes \citep{paluszny2016direct}. We have compared Akaike information criterion \citep{akaike1974new} for both alternatives, which is a common method of model choice \citep{burnham2002practical} and is in our case equivalent to comparing maximum likelihoods in both models (Table \ref{tab::3}) because both models have the same number of parameters. Only for the coarse-grained matrix in '3437-6' is the log-normal distribution slightly more likely. However, as this sample is the smallest, and since the double difference between the likelihoods is also small (only equalled to $\chi_{0.49,2}
^2$ in notation of Eq. \ref{eq::ci}), we have chosen the Weibull distribution for all samples. This also corresponds the observed grain fractioning by grain blocking, which is connected with recrystallisation. For the fine-grained sample, the correspondence between the actual profile size histograms and the densities with ML-estimated parameters is presented in Fig. \ref{fig::5}. The Weibull distribution is indeed preferable, which would be less evident if only non-parametric study were conducted.

 Thus, we assumed the Weibull distribution for all samples. Table \ref{tab::3} presents point and interval estimates for the diameter in the samples, via the range of methods. The typical shape of the two-dimensional confidence ranges for parameters $\lambda$ and $k$ is presented in Fig. \ref{fig::2} for sample 3434-7. The approximation with $m=15$ summands was used for all samples and subsamples. 
\begin{table}[h!]
\centering
\normalsize\begin{tabulary}{1.0\textwidth}{CCCCCC}

Method of estimation & 3437-6 layer      & 3437-6 matrix       & 3434-7             & 3438-3            & 3438-3 subsample  \\ \hline \hline
\multicolumn{2}{l} \mbox{Assumed} Weibull distribution, $m=15$ \\ \hline
ML-1: mean~diameter               & \specialcell{ 3.48 \\(2.82, 4.16)} & \specialcell{ 8.82 \\(5.96, 11.51)}  & \specialcell{ 6.71 \\(4.85, 8.56)}  & \specialcell{ 4.09 \\(3.74, 4.44)} & \specialcell{ 3.71 \\(2.84, 4.56)} \\ \hline
ML-1: $max (log L)$     & -161.2             & -48.77                & -154.58               & -531.65              & -76.7   \\ \hline
ML-2 : mean diameter              & \specialcell{ 3.45 \\(2.85, 4.08)} & \specialcell{ 8.62 \\(5.86,11.41)}   & \specialcell{ 6.58 \\(5.28, 7.98)}  & \specialcell{ 4.09 \\(3.77, 4.41)} & \specialcell{ 3.67 \\(2.93, 4.43)} \\ \hline
ML-3  : mean diameter             & \specialcell{ 3.39 \\(2.79, 4.01)} & \specialcell{ 9.68 \\(6.93, 12.47)}  & \specialcell{ 6.24 \\(4.94, 7.64)}  & \specialcell{ 4.10 \\(3.78, 4.41)} & \specialcell{ 3.69 \\(2.95, 4.45)} \\ \hline
ML weighted             & \specialcell{ 3.42 \\(2.7, 4.1)} & \specialcell{ 9.87 \\(6.9, 12.4)}  & \specialcell{ 6.45 \\(4.9, 8.2)}  & \specialcell{ 4.09 \\(3.8, 4.4)} & \specialcell{ 3.66 \\(2.9, 4.5)} \\ \hline
ML censored : mean diameter         & \specialcell{ 3.61 \\(2.91, 4.28)} & \specialcell{ 11.49 \\(7.51, 15.53)} & \specialcell{ 6.81 \\(4.96, 8.66)}  & \specialcell{ 4.22 \\(3.90, 4.54)} & \specialcell{ 4.38 \\(3.49, 5.27)} \\ \hline
MoM-1   : mean diameter           & \specialcell{ 3.42 \\(2.36, 6.29)} & \specialcell{ 9.23 \\(7.99, 12.68)}  & \specialcell{ 6.67 \\(4.95, 10.78)} & \specialcell{ 4.07 \\(2.69, 7.67)} & \specialcell{ 3.63 \\(2.27, 7.10)} \\ \hline
MoM-2  : mean diameter            & \specialcell{ 3.36 \\(2.30, 6.23)} & \specialcell{ 10.23 \\(8.98,13.68)}  & \specialcell{ 6.38 \\(4.65, 10.48)} & \specialcell{ 4.12 \\(2.73, 7.72)} & \specialcell{ 3.65 \\(2.29, 7.11)} \\ \hline
MDE   : mean diameter               & 3.28              & 8.79                & 6.62               & 3.94              & 3.03              \\ \hline \hline
\multicolumn{2}{l} \mbox{Assumed} log-normal distribution, $m=15$ \\ \hline
ML-1: mean~diameter               & \specialcell{ 3.71 \\(3.16, 4.30)} & \specialcell{ 9.23 \\(7.32, 11.29)}  & \specialcell{ 7.63 \\(5.60, 9.59)}  & \specialcell{ 4.36 \\(4.10, 4.65)} & \specialcell{ 3.75 \\(3.01, 4.52)} \\ \hline
ML-1: $max (log L)$     & -163.41             & -48.09               & -154.25               & -537.08              & -78.00 
\\ \hline \hline

\multicolumn{2}{l} \mbox{Assumed} Weibull distribution, $m=2$ \\ \hline
ML-1: mean~diameter               & \specialcell{ 3.52 \\(2.84, 4.19)} & \specialcell{ 8.39 \\(5.90, 10.78)}  & \specialcell{ 4.06 \\(1.86, 15.31)}  & \specialcell{ 4.01 \\(3.68, 4.34)} & \specialcell{ 3.57 \\(2.73, 4.44)} \\ \hline
ML-1: $max (log L)$     & -163.37             & -48.92               & -151.47               & -541.88              & -79.07 
\\ \hline \hline

Number of profiles   & 81, 85, 12        & 18, 19, 20          & 54, 69, 25         & 271, 301, 78      & 41, 45, 31       

\end{tabulary}
\caption{\label{tab::3}Results for the real samples: estimates and 95 \% interval estimates for mean diameter, along with likelihoods for the models. Abbreviations for the methods --- ML-1: ML, confidence intervals by comparing $2(log(L)-log(\hat{L_0}))$ with its critical value; ML-2: ML, random sampling, bootstrap  intervals; ML-3: ML, all profiles inside the section boundaries are used,  bootstrap  intervals; ML censored: maximising censored likelihood, all profiles are used, confidence intervals by comparing $2log(L-log(\hat{L_0}))$ with its critical value; ML weighted: maximising weighted (Eq. \ref{eq::w}) likelihood, all profiles inside the section boundaries are used, bootstrap  intervals;   MoM-1: ML, random sampling, bootstrap intervals; MoM-2: ML, all profiles inside the section boundaries are used,  bootstrap intervals. 
}
\end{table}

 For the smallest sample -- the '3437 matrix' -- the MoM gives narrower confidence intervals than ML; in other cases, ML is either better or approximately as precise as MoM. We especially note that, although the confidence ranges constructed by comparing the likelihood with its critical value are wider from bootstrap intervals, they guarantee the specified or larger coverage, while bootstrap intervals do not have this property. MDEs are approximately within the confidence range given by the other methods, but the confidence ranges for MDEs are much wider, as observed for the simulated samples. As compared to ordinary likelihood, including all inner profile measurements with bias correction by bootstrapping, neither censored nor weighted estimators improve the precision.
 
 The lower portion of Table \ref{tab::3} and Table \ref{536541} compare  the assumption-based results with the results for the alternative models. The main feature of the wrong choice of log-normal distribution is narrower interval estimates (Table \ref{tab::1}), but qualitatively the estimates remain very close to those in Weibull model. The spherical or spheroidal shape assumption is especially questionable (because profile shapes are fare from circles or ellipses), and to check the spherical shape assumption independently, the mean volume weighted volume of the grains was computed from ML estimates and compared with the unbiased estimator of the mean volume \citep{karlsson1992new}. This check has been done only for samples with $>50$ grains. To roughly determine the imprecision of such estimation, two subsamples were taken from the sample '3438-3' having largest number of profiles. The results are presented in Table \ref{536541}. Another check is that, although the approximation Eq. \ref{eq::4} describes a large class of angular, irregular shapes, it could be that the model with finitely many sides in the polygon that generates the approximate solid will fit the ice grains even better. The ML increases as the approximating shape becomes closer to the sphere in all samples, except the middle-grained sample ‘3434-7’ (Table \ref{tab::1}). This sample was chosen for being presented in this paper because it is non-typical. The larger amount of small sections and the presence of thin, elongated profiles (which may be almost tangential to the particle’s edges) in ‘3434-7’ may be a consequence of polyhedral particle shape, while in other samples the shape is rather irregular. However, the ‘mean diameter’ being so large when estimated with  $m=2$ corresponds to the circumscribed diameter in the model from Fig. \ref{fig::1}. If we choose the inscribed diameter as the ‘effective diameter’, there will be less contrast with the spherical model, as also seen for the mean volume weighted volume cross-check.
\begin{table}[h!]
\centering
\normalsize\begin{tabulary}{1.0\textwidth}{CCCCC}

 & 3437-6 layer        & 3434-7             & 3438-3 left part           & 3438-3 right part  \\ \hline 
ML estimation,
assumed spheres with Weibull distributed diameters& 276 & 3128 &  214 & 207 \\ 
unbiased estimator    & 334             & 2699               & 211                 & 225

\end{tabulary}
\caption{Assumption-based and unbiased assumption-free estimates of mean volume weighted volume
{\label{536541}}%
}
\end{table}

\par\null

\section*{Discussion}

Regarding both bias and lower variance, the ML method is either preferable or similar for estimating mean diameter, as compared to MoM, for the two particle size distributions common in petrography (Weibull and log-normal), as well as the positive normal,  and for sample sizes ranging from 50 - 70, where the estimates are actually informative. The simulations reveal this for samples like ours in any modification of ML and MoM. The only exception may be the coarsest-grained fabric, in which sample size is 18 - 19 and MoM results in narrower bootstrap ranges.

For the Weibull, log-normal and positive normal distributions, a comparable small-sample alternative to ML is MoM, but not MDE.  Not only is MDE imprecise for small samples, but it is also computationally slow.  Unfortunately, MoM is not generally applied in the petrographical context, although both the grounds for constructing estimates with this method \citep[p. 312]{casella2002statistical} and the moments of the profile distributions \citep[p. 37]{baddeley2004stereology} are well-known. For Weibull and log-normal distributions, which are usual in petrology, MoM does not require any complicated software, and the estimates are computed almost immediately. The main disadvantage to using MoM  when samples are very small, however, is the unclear variance. It can be determined by bootstrap, but the precision of the bootstrap estimates and the coverage of the corresponding intervals is unknown without further, thorough study. For relatively small samples, the estimates by both MoM and ML remain imprecise, but non-parametric analogues provide even larger estimate variance, of general reasons \citep[e.g.][p.692]{larsen2013introduction} and as observed in our simulations (Table \ref{tab::1.5}).

Although ML point estimates have similar properties to MoM estimates, ML is also applicable for direct interval estimation, hypothesis testing and model choice, in addition to point estimation. The nearly constant critical value of $2 (log L - log \hat{L_0})$ makes it possible to construct a confidence interval with known coverage. The flexibility of the method allows researchers to consider sampling issues, like sampling all the profiles in a small section by involving the weights. However, weighted sampling does not improve the estimates for samples like ours, although it is a straightforward technique for small samples (by analogy with \citep[e.g.][]{miles1978sampling}). Neither is the precision increased by using measurements all the inner profiles with bootstrap bias correction or censored likelihood. Probably, these ways of considering non-random sampling may be more helpful for other type of samples.

These facts became observable because the present approximation Eq. \ref{eq::4} helps overcome the numerical difficulties of using ML. The imprecision caused by choosing about 15 summands in Eq. \ref{eq::4} is negligible compared to the standard deviation of the estimates, and ML runs at least approximately seven to 20 times faster than MDE, which is itself optimised to run approximately six times faster than \verb"MATLAB". When particles are non-spherical, the error caused by the actual non-spherical shape of the particle must be much larger than the error caused by the non-spherical shape of the approximating model solid. Therefore, we suggest using the present approximation with, at most, ten to 15 summands in practical applications. 
 
It is generally accepted that, when studying irregular particles in this way, researchers study only the ‘effective diameter’, while the revealed spatial patterns and trends are shifted and deformed. However, this study reveals  that the likelihood of particle shape increases towards the spherical model in all studied samples. Moreover, for fine-grained samples, the mean volume weighted volume of the spheres distributed with estimated parameters is very close to the unbiased estimates of the mean volume weighted volume. Applying ML gives ground for the old tradition of using spheres to approximate the mineral particles of convex irregular shapes, and the method is applicable to this study’s petrographic samples.

\section*{Acknowledgements}\label{acknowledgements}
The author would like to thank M. Montagnat (Univ. Grenoble Alpes) for providing the ice micrographs, V.Ya. Lipenkov (AARI) for motivating this study and the anonymous referees of the Image Analysis \& Stereology for very thorough reading and many valuable comments.   
The research was financially supported by the Russian Science Foundation, grant 18-17-00110.
\section*{Conflict of interest}

{\label{644442}}

The author declares no competing interests.
\selectlanguage{english}

\begingroup

\bibliographystyle{plainnat}

\bibliography{references}

\endgroup

\end{document}